\batchmode
\documentstyle[11pt,emulateapj]{article}
\begin{document}
\newcommand{\simlt}{\lesssim}
\newcommand{\simgt}{\gtrsim}
\newcommand{\msol}{M_{\sun}}
\newcommand{\kB}{k_B}
\newcommand{\Cfac}{{\cal C}}
\newcommand{\Ag}{{\cal A}_{\rm g}}
\newcommand{\rhon}{\rho_{\rm n}}
\newcommand{\rhonc}{\rho_{\rm{n,c}}}
\newcommand{\sign}{\sigma_{\rm n}}
\newcommand{\signc}{\sigma_{\rm{n,c}}}
\newcommand{\signci}{\sigma_{\rm{n,c0}}}
\newcommand{\gr}{g_r}
\newcommand{\nn}{n_{\rm n}}
\newcommand{\mn}{m_{\rm n}}
\newcommand{\cc}{\rm{cm}^{-3}}
\newcommand{\muc}{\mu_{\rm c}}
\newcommand{\mucp}{\mu_{\rm c}^\prime}
\newcommand{\muci}{\mu_{\rm{c0}}}
\newcommand{\lref}{l_{\rm{ref}}}
\newcommand{\nnci}{n_{\rm{n,c0}}}
\newcommand{\nnc}{n_{\rm{n,c}}}
\newcommand{\xxi}{x_{\rm{i}}}
\newcommand{\xxic}{x_{\rm{i,c}}}
\newcommand{\xxici}{x_{\rm{i,c0}}}
\newcommand{\xg}{x_{\rm{g}}}
\newcommand{\xgc}{x_{\rm{g,c}}}
\newcommand{\xgci}{x_{\rm{g,c0}}}
\newcommand{\xgcf}{x_{\rm{g,cf}}}
\newcommand{\Bzci}{B_{z,\rm{c0}}}
\newcommand{\Bzc}{B_{z,{\rm c}}}
\newcommand{\Bz}{B_{z}}
\newcommand{\Pext}{P_{\rm{ext}}}
\newcommand{\Pextd}{\tilde{P}_{\rm{ext}}}
\newcommand{\nni}{n_{\rm{i}}}
\newcommand{\nnic}{n_{\rm{i,c}}}
\newcommand{\mg}{m_{\rm g}}
\newcommand{\nng}{n_{\rm{g}}}
\newcommand{\nngc}{n_{\rm{g,c}}}
\newcommand{\nngci}{n_{\rm{g,c0}}}
\newcommand{\siggn}{\langle \sigma w \rangle_{\rm{gn}}}
\newcommand{\Myr}{{\rm{Myr}}}
\newcommand{\nncrit}{n_{\rm{n,c,crit}}}
\newcommand{\tcrit}{t_{\rm{crit}}}
\newcommand{\vd}{v_{{}_{\rm D}}}
\newcommand{\Bref}{B_{\rm ref}}
\newcommand{\muG}{\mu{\rm G}}
\newcommand{\vn}{v_{\rm n}}
\newcommand{\vi}{v_{\rm i}}
\newcommand{\vg}{v_{\rm g}}
\newcommand{\ul}{\underline{\hspace{40pt}}}
\newcommand{\tcore}{t_{\rm{core}}}
\newcommand{\tff}{\tau_{\rm{ff}}}
\newcommand{\tffc}{\tau_{\rm{ff,c}}}
\newcommand{\tffci}{\tau_{\rm{ff,c0}}}
\newcommand{\tgr}{\tau_{\rm{gr}}}
\newcommand{\tgrc}{\tau_{\rm{gr,c}}}
\newcommand{\tgrci}{\tau_{\rm{gr,c0}}}
\newcommand{\tni}{\tau_{\rm{ni}}}
\newcommand{\tng}{\tau_{\rm{ng}}}
\newcommand{\tngc}{\tau_{\rm{ng,c}}}
\newcommand{\tngci}{\tau_{\rm{ng,c0}}}
\newcommand{\Deltag}{\Delta_{\rm g}}
\newcommand{\Deltagc}{\Delta_{\rm{g,c}}}
\newcommand{\Deltagci}{\Delta_{\rm{g,c0}}}
\newcommand{\tnic}{\tau_{\rm{ni,c}}}
\newcommand{\tnici}{\tau_{\rm{ni,c0}}}
\newcommand{\tad}{\tau_{{}_{\rm{AD}}}}
\newcommand{\tadc}{\tau_{{}_{\rm{AD,c}}}}
\newcommand{\tadci}{\tau_{{}_{\rm{AD,c0}}}}
\newcommand{\tphic}{\tau_{{}_{\Phi,\rm{c}}}}
\newcommand{\tphici}{\tau_{{}_{\Phi,\rm{c0}}}}
\newcommand{\PhiB}{\Phi_B}
\newcommand{\PhiBci}{\Phi_{B,\rm{c0}}}
\newcommand{\rM}{r_{{}_{M}}}
\newcommand{\rprime}{r^{\prime}}
\newcommand{\Htwo}{\rm{H_2}}
\newcommand{\sigin}{\langle \sigma w \rangle_{\rm{in}}}
\newcommand{\Ageo}{{\sl A}_{\rm{geo}}}
\submitted{To appear in The Astrophysical Journal, vol. 547 \#1, 20 Jan 2001}
\title{ON THE TIMESCALE FOR THE FORMATION OF PROTOSTELLAR CORES IN
MAGNETIC INTERSTELLAR CLOUDS}
\author{Glenn E. Ciolek\altaffilmark{1} and Shantanu Basu\altaffilmark{2}}
\altaffiltext{1}{New York Center for Studies on the Origins of Life (NSCORT),
and Department of Physics, Applied Physics, and Astronomy,
Rensselaer Polytechnic Institute, 110 8th Street, Troy, NY 12180;
cioleg@rpi.edu.}
\altaffiltext{2}{Department of Physics and Astronomy,
University of Western Ontario, London, Ontario N6A 3K7, Canada; 
basu@astro.uwo.ca.}
\slugcomment{To appear in The Astrophysical Journal, vol. 547 \#1, 20 Jan 2001}
\righthead{G. E. Ciolek and S. Basu}
\lefthead{Timescale for Protostellar Core Formation in Magnetic Clouds}
\begin{abstract}
We revisit the problem of the formation of dense protostellar cores
due to ambipolar
diffusion within magnetically supported molecular clouds, and derive
an analytical expression for the core formation timescale. The
resulting expression is similar to the canonical expression $\simeq
\tff^2/\tni \sim 10 \tff$ (where $\tff$ is the free-fall time and $\tni$ is
the neutral-ion collision time), except that it is multiplied by a
numerical factor
$\Cfac(\muci)$, where $\muci$ is the initial central mass-to-flux
ratio normalized to the critical value for gravitational collapse.
$\Cfac(\muci)$ is typically $\sim 1$ in highly subcritical clouds
($\muci \ll 1$), although certain conditions allow $\Cfac(\muci) \gg 1$.
For clouds that are not highly subcritical, $\Cfac(\muci)$ can be much less
than unity, with $\Cfac(\muci) \rightarrow 0$ for
$\muci \rightarrow 1$, significantly reducing the time required to
form a supercritical core. This, along with recent observations of
clouds with mass-to-flux ratios close to the critical
value, may reconcile the results of ambipolar diffusion models with
statistical analyses of cores and YSO's which suggest an evolutionary
timescale $\sim$ 1 Myr for objects of mean density $\sim 10^4 \cc$.
We compare our analytical relation to the
results of numerical simulations, and also discuss the effects of
dust grains on the core formation timescale.

\end{abstract}
\section{Introduction}
Ambipolar diffusion, the drift of neutral matter with respect to
magnetic field and plasma, was put forth by Mestel \& Spitzer (1956)
as a means by which magnetic flux can escape from interstellar clouds
and thereby induce overall collapse and fragmentation. To explain the
well-known inefficiency
of star formation, Mouschovias (1976, 1977, 1978) proposed instead that,
rather than reduce the flux of a cloud as a whole, ambipolar diffusion
gravitationally redistributes matter within the central flux tubes of a
magnetically supported cloud.
He further suggested (Mouschovias 1979, 1982) that the formation of
cores would occur on the ambipolar diffusion timescale
$\tad \simeq \tff^2/\tni$, where $\tff$ and $\tni$ are respectively,
the free-fall and neutral-ion collision times (discussion is also
provided in Mouschovias 1987; McKee et al.  1993; Galli \& Shu 1993;
and Mouschovias \& Ciolek 1999). For typical ion abundances
within molecular clouds (e.g., Gu\'{e}lin, Langer, \& Wilson  1982;
Langer 1985; McKee 1989; Ciolek \& Mouschovias 1994, 1998),
one finds $\tad \sim 10 \, \tff$ over a wide range of densities.

Numerical simulations in
axisymmetric (Fiedler \& Mouschovias 1993) and disk-like (Ciolek \&
Mouschovias 1994 [CM94], 1995; Basu \& Mouschovias 1994 [BM94],
1995a, b) model clouds confirmed the theoretical scenario described above.
No magnetic flux leaked out of these model clouds, while in the
interior flux tubes supercritical cores formed through ambipolar
diffusion; core-envelope separation ensued, in which dense cores
collapsed dynamically while embedded in massive, magnetically supported
(subcritical) envelopes. For highly subcritical model clouds
(i.e., with initial central mass-to-flux ratio less than
$1/e$ times the critical value), cores did
indeed form (i.e., the central mass-to-flux ratio reached the critical
value) on a timescale $\tcore \simeq$ the initial central
flux-loss time $\tphici = \tadci/2$ (e.g., CM94, \S~3; BM94, \S~4).
However, model clouds that had initial central mass-to-flux ratios
$(M/\PhiB)_{\rm{c0}}$ that were less subcritical formed in a time
$\tcore$ significantly less than $\tphici$ (e.g., model 5 of Basu \&
Mouschovias 1995b). This was also true for the recent model presented
by Ciolek \& Basu (2000; CB00), which successfully reproduced
observations of the mass and velocity structure within the L1544
prestellar core. For this model $\muci = 0.8$, where
$\muci = (M/\PhiB)_{\rm{c0}}/(M/\PhiB)_{\rm{crit}}$ is the initial
central mass-to-flux ratio in units of the critical value for
gravitational collapse, and $\tcore$ was found to be
$\simeq 0.2 \tphici$. This suggests that $\tcore$ is
dependent on the value of $\muci$. In this paper we derive an
analytical relation for $\tcore(\muci)$ that is accurate for {\em all}
subcritical clouds (i.e., $\muci \leq 1$).

In addition to its general theoretical interest,
this investigation is particularly timely, given that a recent
compilation of magnetic field strength data (Crutcher 1999) shows that
many clouds have mass-to-flux ratios close to the critical
value, and some recent statistical studies of ages and lifetimes of
protostellar cores and young stellar objects which seem to indicate that
star formation can occur on a timescale $\sim 1$ Myr (e.g., Lee \& Myers
1999; Jijina, Myers, \& Adams 1999) within objects of mean density
$\sim 10^4 \cc$; that is, it occurs in $\sim$ few $\tff$ rather than
$\sim 10 \, \tff$ at these densities. Our general results in this paper
(as well as the specific case presented in CB00) show that these
inferred lifetimes are compatible with the observed regions being
slightly subcritical and evolving due to ambipolar diffusion, in
addition to the possibility that they are already somewhat supercritical
and evolving dynamically.
\section{Derivation of the Core Formation Timescale}
The fundamental equations we use were developed for disk-like model
molecular clouds (using a cylindrical polar coordinate system
$r,\phi,z$), and were presented in \S~3 of Ciolek \& Mouschovias
(1993; CM93) and \S~2 of BM94; the results we derive here can be
generalized to other geometries in a straightforward manner. Model
clouds are isothermal, embedded in an external medium with constant
pressure $\Pext$, and contain neutral molecules and atoms ($\rm{H_2}$
with a cosmic abundance of He), trace amounts of ions (atomic and
molecular), electrons, and charged and neutral dust grains.
The clouds are self-gravitating and magnetically supported,
with $\muci \leq 1$.

As in Ciolek \& Mouschovias (1996; CM96) we start by considering 
the motion of a neutral fluid element in a flux tube centered on the
axis of symmetry ($r=0$) containing mass $M$ and magnetic flux
$\PhiB$ and intersecting the equatorial plane at a Lagrangian radius 
$\rM(t)$. In a frame moving with the neutrals,
\begin{eqnarray}
\label{dmdteqa}
\frac{d}{dt}\left(\frac{M}{\PhiB}\right)
&=&-\frac{M}{\PhiB^2}\frac{d \PhiB}{dt} 
= -2\pi \frac{M}{\PhiB^2} \frac{d}{dt}
\int_{0}^{\rM} d\rprime \rprime \Bz(\rprime) \nonumber \\
&=& 2 \pi \frac{M}{\PhiB^2} \rM \Bz(\rM) \vd(\rM),
\end{eqnarray}
where $\Bz$ is the magnetic field and $\vd \equiv \vi - \vn$
is the radial ion-neutral drift velocity; $\vn = d\rM/dt$ is the neutral
velocity and $\vi$ is the ion velocity. In deriving this equation
we have used the magnetic induction equation
$\partial \Bz/\partial t = -(1/r)\partial(r \Bz \vi)/\partial r$. The
induction equation reflects the freezing of magnetic flux in the ions,
which is valid for the typical density regime in which cores form
($\nn \simlt 10^5~\cc$; see \S~3.1.2 of CM93).
Core formation takes place in the innermost flux tubes of a cloud,
and so we consider equation (\ref{dmdteqa}) in the limit
$\rM \rightarrow 0$. The numerical simulations cited in \S~1 found that
$\Bz$ was spatially uniform in this limit, therefore we may take
$\PhiB \simeq \pi \Bz \rM^2$.
Inserting this relation into equation (\ref{dmdteqa}), and dividing both
sides by $(M/\PhiB)_{\rm{crit}}$, yields
\begin{equation}
\label{dmucdteq}
\frac{1}{\muc} \frac{d\muc}{dt}
= \frac{2 \vd}{\rM} = \frac{1}{\tphic} . 
\end{equation}
Ambipolar diffusion in the central flux tubes results in $\vd >0$,
which means that plasma and magnetic flux are ``left behind" as neutrals
drift inward toward the symmetry axis (due to self-gravity, see below),
and equation (\ref{dmucdteq}) shows that the dimensionless central
mass-to-flux ratio $\muc$ will increase with an instantaneous
$e$-folding timescale $\tphic \equiv \rM/2\vd(\rM)$. Given the
definition of the ambipolar diffusion timescale $\tad \equiv r/\vd$
(e.g., Spitzer 1978, \S~13.3e; Mouschovias 1979), it immediately follows
that $\tphic = \tadc/2$. The behavior of $\tphic$ during the evolution
of various cloud models is depicted in figures $2d$ and $4d$ of CM94,
and figures $2c$ and $6c$ of BM94; examination of these figures
indicates that $\tphic$ normally {\em decreases} as a cloud evolves
(i.e., as $\muc$ increases with time).

A supercritical core exists for $\muc \geq 1$. Core formation occurs
at a time $\tcore$ when $\muc(\tcore)=1$. Integrating equation
(\ref{dmucdteq}) from $t=0$ to $t=\tcore$ gives
\begin{equation}
\label{tcoreeqa}
\tcore = \int_{\muci}^{1} \frac{d\mucp}{\mucp} \tphic(\mucp) .
\end{equation}
Thus, we may write
\begin{mathletters}
\begin{equation}
\label{tcoreeqb}
\tcore = \Cfac(\muci) \tphici,
\end{equation}
where
\begin{equation}
\label{Cfacdefeq}
\Cfac(\muci) = \int_{\muci}^{1} \frac{d\mucp}{\mucp}
\frac{\tphic(\mucp)}{\tphici} ,
\end{equation}
\end{mathletters}
and $\tphici \equiv \tphic(\muci)$.
The factor $\Cfac(\muci)$ defined in equations (\ref{tcoreeqb}) and
(\ref{Cfacdefeq}) is the
{\em initial flux constant} of a cloud. In a sense, it is a measure of
the ``distance a cloud has to go" in its central mass-to-flux ratio to
attain the critical state $\muc=1$, given a particular initial value
$\muci$. The ``closer" $\muci$ is to unity, the smaller we expect
$\Cfac(\muci)$ to be. Since $\tphic(\muc) \leq \tphici$ for
$\muc \geq \muci$ (as noted above), equation (\ref{Cfacdefeq})
provides the limiting relation
\begin{equation}
\label{tcorelimiteq}
\Cfac(\muci) \leq \ln\left(\frac{1}{\muci}\right).
\end{equation}
Thus, $\Cfac(\muci) \rightarrow 0$ as $\muci \rightarrow 1$.

To proceed further, we need an expression for $\tphic(\muc)$ to
evaluate the integral in equations (\ref{tcoreeqa}) and (\ref{Cfacdefeq}).
This can be
obtained from the force equation (per unit area) for the neutrals,
which is given by equations (28a)-(28c) and (50) of CM93. In the next
subsection we first consider models that account only for neutral-ion
collisions and neglect the friction due to grains. The effect of
neutral-grain collisions are then discussed in the following subsection.
\subsection{Clouds With Only Neutral-Ion Collisions}
The numerical simulations referred to in \S~1 generally found
that the evolution during the core formation epoch is quasistatic,
with little or no acceleration of the neutral particles. Ignoring the
acceleration of the neutrals, as well as thermal pressure stresses
(which are also negligible at these densities and lengthscales),
the evolution of the neutrals is determined by a balance between
gravitational and collisional forces: 
\begin{equation}
\label{forceeqa}
0 \simeq \sign \gr + \frac{\sign}{\tni} \vd,
\end{equation}
where $\sign$ is the mass column density of the neutrals, $\gr$ is the
radial gravitational field, and $\tni$ is the neutral-ion collision
(momentum exchange) time. Solving this equation, we find
$\vd = \tni |\gr|$. Whence, again in the limit $\rM \rightarrow 0$,
\begin{equation}
\label{tphieqa}
\tphic = \frac{\rM}{2 \vd(\rM)}
= \frac{1}{2}\frac{\rM}{\left(|\gr| \tni\right)_{\rM}}
= \frac{1}{2} \left(\frac{\tgr^2}{\tni}\right)_{\rm c}
\simeq \frac{1}{2}\left(\frac{\tff^2}{\tni}\right)_{\rm c}  ,
\end{equation}
where $\tgr \equiv (r/|\gr|)^{1/2}$ is the gravitational contraction
timescale (Ciolek \& K\"onigl 1998), which is $\simeq \tff$.
(Note: in both CM94 and BM94, $\tgr$ is referred to as the dynamical
timescale $\tau_{\rm{dyn}}$.) 

For $\tgr$ we may write
\begin{equation}
\label{tgreq}
\tgr = \left(\frac{\Ageo}{G \mn \nn}\right)^{1/2} ,
\end{equation}
where $G$ is the gravitational constant, $\mn$ is the mean mass
of a neutral particle (= 2.33 amu for an $\Htwo$ gas with a cosmic
helium abundance), and $\Ageo$ is a numerical constant that is solely
dependent on the assumed geometry of a model cloud. For most geometries,
$\Ageo^{1/2} \approx 1$.

The neutral-ion collision time is expressed as
\begin{equation}
\label{tnieq}
\tni = 1.4 \left(1 + \frac{m_{{}_{\Htwo}}}{m_{\rm i}} \right)
\frac{1}{\nni \sigin} ,
\end{equation}
where $m_{{}_{\Htwo}}$ and $m_{\rm i}$ are the masses of $\Htwo$ molecules
and ions, respectively, $\nni$ is the ion density, and $\sigin$ is the
ion-neutral collisional rate, which is
$\simeq 1.7 \times 10^{-9}~\rm{cm}^3~{\rm s}^{-1}$ for
$\Htwo$-$\rm{Na}^{+}$ and $\Htwo$-$\rm{HCO}^{+}$ collisions
(McDaniel \& Mason 1973).
The factor 1.4 in equation (\ref{tnieq}) accounts for the fact
that the inertial effect of helium is neglected in the 
neutral-ion collisional force; for further discussion on this
point we refer the reader to \S~2.1 of Mouschovias \& Ciolek (1999).

Finally, we parameterize the central ion density through the relation
$\nnic \propto \nnc^k$, where, in this analysis, we assume that
$k$ is a constant. Strictly speaking, $k = k(\nnc)$, with
$0.5 \leq k \simlt 1$ for densities $\nnc \simlt 10^5~\cc$, as shown
in numerical simulations (CM94) and analytical studies of the
effect of ambipolar diffusion on ion abundances in contracting cores
(Ciolek \& Mouschovias 1998). However, as we shall see below, our
neglect of the dependence of $k$ on the density leads to only a minor
error in the calculation of the integral in equation (\ref{Cfacdefeq}).
Combining this relation with equations (\ref{tphieqa}), (\ref{tgreq}),
and (\ref{tnieq}), and using our parameterization for $\nnic(\nnc)$
discussed above, yields
\begin{equation}
\label{tphieqb}
\frac{\tphic}{\tphici}= \left(\frac{\nnci}{\nnc}\right)^{1-k} .
\end{equation}

Our remaining task is to determine $\nnc(\muc)$. In the disk model
clouds, the mass density $\rhon(r)$ [$=\mn \nn(r)$] is related to the
column density $\sign(r)$ in a magnetic flux tube through balance of
thermal-pressure forces and self-gravity along the $z$-direction
(i.e., along the direction of the magnetic field):
\begin{equation}
\label{rhoneq}
\rhon C^2 = \Pext + \frac{\pi}{2} G \sign^2 ~,
\end{equation}
$C$ being the isothermal speed of sound (for a derivation, see
\S~3.1.1 of CM93). It then follows that
\begin{equation}
\label{densrateq}
\frac{\nnci}{\nnc} =
\left[\frac{\Pextd + 1}{\Pextd + \left(\signc/\signci\right)^2}\right]~,
\end{equation}
where $\Pextd \equiv 2 \Pext/\pi G \signci^2$ is the dimensionless
external pressure, in units of the initial self-gravitational stress in
the central flux tube. Now, in the limit $\rM \rightarrow 0$,
$(M/\PhiB)_{\rm c} = \signc/\Bzc$, and, therefore,
\begin{equation}
\label{muapproxeq}
\frac{\muc}{\muci}
= \frac{\left(M/\PhiB\right)_{\rm c}}{\left(M/\PhiB\right)_{\rm{c0}}}
= \left(\frac{\signc}{\signci}\right) \left(\frac{\Bzci}{\Bzc}\right)
\approx \frac{\signc}{\signci} ~.
\end{equation}
In the last equality in equation (\ref{muapproxeq}) we have used
the fact that the central magnetic field strength $\Bzc$ 
changes very little during the core formation epoch
(Fiedler \& Mouschovias 1993; CM94; BM94). That is, during this stage
of evolution, ambipolar diffusion allows magnetic field lines to remain
essentially ``fixed in place" (Mouschovias 1978, 1979).
Inserting equation (\ref{muapproxeq}) into (\ref{densrateq}), and then
substituting into (\ref{tphieqb}) and (\ref{Cfacdefeq}), we have
\begin{equation}
\label{Cfaceqa}
\Cfac(\muci)= \int_{1}^{\frac{1}{\muci}} \frac{dy}{y} \left(\frac{\Pextd + 1}{\Pextd + y^2}\right)^{1-k}, 
\end{equation}
where $y \equiv \muc/\muci$.
For a given value of $\Pextd$, $\Cfac(\muci)$ can be obtained by 
integration of equation (\ref{Cfaceqa}).
Note that for $k=1$, equation (\ref{Cfaceqa}) readily yields
\begin{equation}
\label{Cfaceqb}
\Cfac(\muci) = \ln\left(\frac{1}{\muci}\right),
\end{equation}
regardless of the value of $\Pextd$.
Similarly, for the case $k=1/2$, the integral (\ref{Cfaceqa})
has the solution
\begin{equation}
\label{konehalfeq}
\hspace{-2em}\Cfac(\muci) = \left(\frac{\Pextd + 1}{\Pextd}\right)^{1/2}
\ln \left[\frac{ \Pextd^{1/2} + \left(\Pextd + 1 \right)^{1/2}}
{\Pextd^{1/2}\muci + \left(\Pextd\muci^2 + 1 \right)^{1/2}}\right].
\end{equation}

The previously published model clouds cited in \S~1 typically adopted a
value $\Pextd = 0.1$, appropriate for an isolated, gravitationally
bound cloud. For such models the integral (\ref{Cfaceqa}) may be
solved by taking the limit $\Pextd \rightarrow 0$, yielding
\begin{equation}
\label{Cfaceqc}
\Cfac(\muci) \simeq \frac{1 - \muci^{2(1-k)}}{2\left(1-k\right)} 
\hspace{3em}{\rm{for}}~k < 1
\end{equation}
(for $k=1/2$, this expression is also easily obtained from eq.
[\ref{konehalfeq}] in the same limit),
and for $k=1$, $\Cfac(\muci)$ is given by (\ref{Cfaceqb}) above.
Figure 1 shows $\Cfac(\muci)$ as a function of $\muci$, for models
with $\Pextd = 0.1$ and $k=0.5$, 0.6, 0.75, and 1.0, respectively. 
The {\it solid} lines are the solutions given by equations
(\ref{Cfaceqb}) and (\ref{Cfaceqc}), which assume $\Pextd = 0$.
The curves with {\it crosses}, for the models with $k=0.5$, 0.6, and 0.75,
represent the solution for $\Pextd = 0.1$, obtained from equation
(\ref{Cfaceqa}) or direct numerical
integration. Examination of this figure reveals that the analytical
approximation (\ref{Cfaceqc}) for these small $\Pextd$ models is in
excellent agreement with the exact integration.

Also shown in Figure 1 are the results for
$\Cfac(\muci) = \tcore/\tphici$ as found in some of our earlier
published core formation and collapse numerical simulations: model A of
CM94 ($\Pextd = 0.1$, $\muci = 0.26$, $k \simeq 0.65$ --- see Fig. $2c$
of CM94), model 2 of BM94 ($\Pextd = 0.1$, $\muci =0.324$, $k=0.5$),
model 5 of BM95b ($\Pextd =0.1$, $\muci =0.5$, $k=0.5$), 
and the L1544 model of CB00 ($\Pextd =0.1$, $\muci=0.8$,
$k \simeq 0.7$).
Comparison of the values found in the numerical
simulations with those derived from our expressions for
$\Cfac(\muci)$ (eqs. [\ref{Cfaceqa}]-[\ref{Cfaceqc}]) also show
excellent agreement.

Notice that, for highly subcritical clouds ($\muci \ll 1$),
$\Cfac(\muci) \sim 1$ for most realistic values of $k$. Hence, for
these clouds, $\tcore \approx \tphici$, as originally argued by
Mouschovias (1979, 1982).
Models with $k \simeq 1$ and $\muci \ll 1$ are an exception:
for these models $\tphici$ does not decrease with increasing density,
since the degree of ionization $\xxic = \nnic/\nnc \propto \nnc^{k-1}$
in the central flux tubes remains essentially constant as the cloud
evolves. Under these conditions, $\tphici$ is a true $e$-folding
timescale.

However, $\Cfac(\muci) \ll 1$ as $\muci$ approaches unity, for all
values of $k$. This region is of particular interest, given the recent
OH Zeeman observations of magnetic field strengths in dense molecular
cloud gas by Crutcher (1999). The estimated mass-to-flux ratios are
close enough to the critical value that a clear observational distinction
between subcritical and supercritical cores cannot be made. However, when 
the likely effects of random
orientations of cloud axes and magnetic field projections on the plane
of the sky are accounted for, Crutcher's inferred values are consistent
with $0.5 \simlt \muc \simlt 1.5$ (e.g., see Table 1 of Shu et al. 1999).
Although his sample undoubtedly contains some cores that 
have already
become supercritical (and therefore are in the early stages of
collapse), these values may also be indicative of $\muci$
for typical star-forming molecular clouds. In this case, clouds may
often be subcritical (and magnetically supported), but
only moderately so.
\footnote{That clouds may have mass-to-flux ratios near the critical
value is consistent with the well-known linewidth-size relation for
molecular clouds (e.g., Larson 1981; Myers 1983). For instance,
Mouschovias \& Psaltis (1995) analyzed the data for the linewidths in 14
different clouds and showed that the ubiquitous relation
$\Delta v \propto R^{1/2}$ (where $\Delta v$ is the linewidth and $R$
the radial extent of a cloud) could arise from turbulent or nonthermal
linewidths resulting from hydromagnetic waves in clouds with magnetic
field strengths comparable to the critical value for collapse.}
From equations (\ref{Cfaceqb}) - (\ref{Cfaceqc}) it follows that,
for $(1-\muci) \ll 1$, $\Cfac(\muci) \simeq 1 - \muci$.
Hence, $\Cfac(\muci) \ll 1$ will be expected to occur
for clouds with $\muci$ values in the range suggested by the Crutcher
(1999) data. {\em This means that clouds with these values of $\muci$
will evolve and form cores by ambipolar diffusion on a timescale much
less than $\tphici$.}

Let us evaluate the magnitude of this effect numerically. From
equations (\ref{tphieqa}), (\ref{tgreq}), and (\ref{tnieq})
we find
\begin{equation}
\label{tphicivaleq}
\tphici = 7.0 \times 10^6 \Ageo
\left(\frac{\xxici}{10^{-7}}\right)~\rm{yr},
\end{equation}
where we have normalized $\xxici$ to a value typical for cosmic-ray 
ionized gas at density $n \simeq 10^4 \cc$.
Since, according to our results displayed in Figure 1,
$\Cfac(\muci) \simlt 0.5$ for clouds with $\muci \simgt 0.5$, we
conclude from equations (\ref{tcoreeqb}) and (\ref{tphicivaleq}) that
$\tcore \simlt 4$ Myr for moderately subcritical clouds. This effect
had already been seen in the L1544
model of CB00, which had $\muci =0.8$, resulting in $\Cfac(\muci)=0.2$,
and $\tcore = 1.3$ Myr (see their \S~3). {\em Thus, recent statistical
analyses suggesting that star formation occurs in some regions on
timescales $\sim$ 1 Myr may be explained by ambipolar diffusion models
and observations of subcritical clouds with central mass-to-flux ratios slightly
below the critical value for collapse.} We also note that the much longer 
lifetimes of a few $\times \, 10^7$ yr for giant molecular clouds as 
a whole can be accounted for by their lower mean density, yielding a 
higher ionization fraction $\xxi$, particularly when ionization due to
background ultraviolet starlight is accounted for (e.g., Ciolek \&
Mouschovias 1995). The higher values of $\xxi$ can greatly extend the
lifetime of lower density regions even if the mass-to-flux ratio is
close to critical.

In the limit $\Pextd \gg 1$, corresponding to model clouds situated in a
high-pressure external medium, appropriate for cloud fragments or
sheets located in a dense or massive cluster or cloud complex, or in a
particularly hot medium or intense radiation field (perhaps suitable
for proplyds), the integral in equation (\ref{Cfaceqa}) is found in the
limit $\Pextd \rightarrow \infty$ to be the same as that given by
equation (\ref{Cfaceqb}). For this case, clouds with $\muci < 1/e$
have $\tcore > \tphici$ since $\tphici$ is again an $e$-folding
timescale. This occurs in this limit because the mass density $\rhon$
given by equation (\ref{rhoneq}) is insensitive to the amount of mass
within a flux tube (given by the column density $\sign$). This is seen
from equation (\ref{densrateq}), which implies that the fractional
variation of the density $\delta \nnc/\nnc$ is related to the
fractional variation in column density $\delta \signc/\signc$ by
\begin{eqnarray}
\label{denvarianceq}
\frac{\delta \nnc}{\nnc} &=&
2 \frac{\left(\signc/\signci\right)\left(\delta \signc/\signci\right)}
{\left(\signc/\signci\right)^2 + \Pextd} \nonumber \\
&\approx& 2 \frac{\left(\muc/\muci\right) \left(\delta \muc/\muci\right)}
{\left(\muc/\muci\right)^2 + \Pextd} ,
\end{eqnarray}
where the last approximate equality is due to equation
(\ref{muapproxeq}). For $\Pext \rightarrow 0$, it follows from 
(\ref{denvarianceq}) that
$\delta \nnc/\nnc \rightarrow 2 \delta \signc/\signc \approx
2 \delta \muc/\muc$,
i.e., the central density increases twice as fast as the column
density and mass-to-flux ratio. In the opposite extreme,
$\Pextd \rightarrow \infty$,
$\delta \nnc/\nnc \rightarrow 2 (\signc/\signci)(\delta \signc/\signci)/\Pextd
\approx 2 (\muc/\muci)(\delta \muc/\muci)/\Pextd \rightarrow 0$.
For these models, $\nnc$ is determined primarily by vertical
confinement (compression) of the disk by the external bounding pressure
$\Pext$, and, as a result, there is little change in the density
as the column density and mass-to-flux ratio increases in the
central flux tubes. Since the density does not increase much in this
case, neither does the degree of ionization and $\tphic$ is effectively
constant as the cloud evolves; this follows from equation
(\ref{tphieqb}), which yields
$\delta \tphic/\tphic = -(1-k)\delta \nnc/\nnc$. (As noted earlier,
$k=1$ also yields $\delta \tphic =0$.) Figure 2 displays $\Cfac(\muci)$
for two model clouds with $\Pextd = 20$; one has $k=0.5$ and the other
$k=1$. The high-pressure analytical result
$\Cfac(\muci) = \ln(1/\muci)$, is a good approximation
for these models. In summary, we have demonstrated that ambipolar
diffusion driven evolution actually occurs less rapidly in regions of
high external pressure, since the density, which is then determined to
a great extent by the bounding pressure, increases less rapidly as the
central mass-to-flux ratio increases.

Finally, before ending this subsection we make an important physical 
distinction. As described above, our new analysis shows that, for clouds
with $\muci \sim 1$, $\Cfac(\muci)  = \tcore/\tphici \ll 1$. Despite
the fact that a core forms at a faster rate under these conditions
(when compared to the uncorrected initial flux-loss timescale), it
should not be considered to be dynamical. Recall
that our results are obtained using a {\em quasistatic} approximation
for the equation of motion of the neutrals during the core formation
epoch, consistent with behavior seen in our detailed numerical
simulations. Therefore, although the timescale for core formation in
this situation is much less than that which would have been predicted
by the ``classical" expression for a core forming due to ambipolar
diffusion (i.e., $\tcore \simeq \tphici$), it should {\em not} be taken
to mean that the mass in the inner flux tubes is undergoing dynamical or
accelerated collapse during this phase of evolution. 
As described in detail in the numerical simulations cited in \S~1,
magnetically-diluted collapse occurs within a supercritical core
only after it forms, leading to star formation in an additional
few $\tff$.
\subsection{Inclusion of Grain Collisional Friction}
We now consider the effect of neutral-grain collisions. Here we restrict
our analysis to standard models with $\Pextd \ll 1$. Detailed
discussion of the effects of grains (charged and neutral) on the
evolution of self-consistent model clouds is provided in CM93, CM94, and
CM95. For typical grain sizes in dense molecular clouds
(radii $a \simgt 10^{-6}~\rm{cm}$) the bulk of the grains have a charge
$-e$, where $e$ is the electronic charge. Moreover, as shown in CM93-95,
rapid inelastic capture of ions and electrons on grains transfers
momentum between charged and neutral grains, which can also result in
effective coupling of the neutral grains to the magnetic field.

We again make use of the fact that the evolution due to ambipolar
diffusion during the core formation epoch is well-approximated as being
quasistatic. The motion of the neutrals is therefore given by
\begin{equation}
\label{forceeqb}
0 \simeq \sign \gr + \frac{\sign}{\tni} \vd
+ \frac{\sign}{\tng}\Deltag \vd.
\end{equation}
The latter term on the RHS of equation (\ref{forceeqb}) is the force
(per unit area) from neutral-grain collisions. The quantity $\tng$ is
the neutral-grain collision time and $\Deltag$
[$ \equiv (\vg - \vn)/\vd$, where $\vg$ is the grain velocity] is the
{\em total magnetic attachment parameter of the grains}, originally
introduced in \S~3.1.2 of CM93.
$\Deltag$ expresses the coupling of charged dust grains to the magnetic
field: grains that are attached to magnetic field lines will have
$\vg = \vi$, and, therefore, $\Deltag = 1$; in the opposite limit of
unattached grains (due to collisions with neutrals), $\vg = \vn$ and
$\Deltag = 0$.

Equation (\ref{forceeqb}) yields
\begin{equation}
\label{vdeqb}
\vd(\rM)= \left(\frac{|\gr| \tni}{1 + (\tni/\tng)\Deltag}\right)_{\rM}.
\end{equation}
As in the preceding section, in the limit $\rM \rightarrow 0$
it follows from equations (\ref{dmucdteq}) and 
(\ref{vdeqb}) that
\begin{equation}
\label{tphiceqb}
\tphic= \frac{1}{2} \left(\frac{\tgr^2}{\tni}\right)_{\rm c}
+ \frac{1}{2} \left(\frac{\tgr^2}{\tng} \Deltag \right)_{\rm c}. 
\end{equation}
In the derivation of $\Cfac(\muci)$, the evaluation of the first term
on the RHS of equation (\ref{tphiceqb}) --- the contribution to the
flux-loss timescale due to the ions alone (which would occur for
$\Deltag = 0$ or $\tni/\tng \ll 1$) --- follows exactly as presented in
\S~2.1 above. We focus now on the second term on the RHS of
equation (\ref{tphiceqb}). To determine its effect on $\Cfac(\muci)$
involves evaluating the integral
\begin{equation}
\label{grintegraleq}
\hspace{-3ex} \int_{\muci}^{1}\frac{d\mucp}{\mucp} \frac{\tgrc^2}{\tngc} \Deltagc 
=\left(\frac{\tgr^2}{\tng}\right)_{\rm{c0}}
\int_{\muci}^{1}\frac{d \mucp}{\mucp}
\left(\frac{\tgrc}{\tgrci}\right)^2 \frac{\tngci}{\tngc} \Deltagc. 
\end{equation} 
Simplification comes about since
\begin{equation}
\label{intsimpeq}
\left(\frac{\tgrc}{\tgrci}\right)^2 \frac{\tngci}{\tngc}
= \frac{\nnci}{\nnc} \frac{\nngc}{\nngci}
= \frac{\xgc}{\xgci} ,
\end{equation}
where $\nng$ is the density of grains and $\xg = \nng/\nn$ is
the relative abundance of grains. In the reduction of equation
(\ref{intsimpeq}), we have used equation (\ref{tgreq}) and the fact
that the neutral-grain collision time
\begin{equation}
\label{tngeq}
\tng = \frac{\mn + \mg}{\mg} \frac{1}{\nng \siggn} ;
\end{equation}
$\mg$ ($\gg \mn$) is the grain mass, and $\siggn$ is the grain-neutral
collision rate, equal to $\pi a^2 (8 \kB T/\pi \mn)^{1/2}$, where $a$ is
the radius of the grains, $\kB$ is the Boltzmann constant, and $T$ is
the gas temperature.

Further progress is made by using the analysis of CM96, who studied the
effect of ambipolar diffusion on dust-to-gas ratio during the formation
of cores. As CM96 described at length, ambipolar diffusion can reduce
the abundance of grains in a core when the inwardly drifting neutrals
``leave behind" the magnetic field lines and the grains that are
attached to them. In particular, they showed that during the core
formation epoch, $\Deltagc d \mucp/\mucp = - d \xgc/\xgc$. Therefore,
when $\Deltagc \neq 0$ (corresponding to some degree of attachment
of grains to the magnetic field), the relative abundance of grains
decreases as the central mass-to-flux ratio increases.
\footnote{This particular result is unique to our ambipolar diffusion
models. It also affects the ion chemistry in contracting cores, as
discussed in CM94 and Ciolek \& Mouschovias (1998).}
Using this relation, along with equation (\ref{intsimpeq}), 
the integral (\ref{grintegraleq}) becomes
\begin{equation}
\label{grintegraleqb}
\int_{\muci}^{1} \frac{d \mucp}{\mucp} \frac{\tgrc^2}{\tngc} \Deltagc
= \left(\frac{\tgr^2}{\tng}\right)_{\rm{c0}} \left(1 - \frac{\xgcf}{\xgci}\right) , 
\end{equation}
where $\xgcf$ is the abundance of grains in the central flux tubes
at time $\tcore$, $\muc=1$. CM96 (eq. [11]) also showed that
\begin{equation}
\label{grfraceq}
\frac{\xgcf}{\xgci} =
\muci \left(\frac{1 + \Ag^2}{1 + \Ag^2\muci^4}\right)^{1/4},
\end{equation}
where $\Ag$ is a dimensionless quantity whose inverse is related to the
strength of the attachment of charged grains to the magnetic field.
Specifically, equations (7), (8), and (9a)-(b) of CM96 yield
\begin{equation}
\label{Ageq}
\Deltagci = \frac{1}{1 + \Ag^2 \muci^4}, 
\end{equation}
and therefore, for $\Ag \rightarrow 0$, $\Deltagci \rightarrow 1$
(grains are frozen to magnetic field lines), while,
conversely for $\Ag \rightarrow \infty$, $\Deltagci \rightarrow 0$
(grains are completely unattached). Numerically,
\begin{eqnarray}
\label{Ageqb}
\Ag &=& 2.32 \times 10^{-2}\left(\frac{\Bzci}{35~\mu{\rm G}}\right)
\left(\frac{a}{10^{-6}~{\rm{cm}}}\right)^2 \nonumber \\
&& \hspace{3em} \times \left(\frac{\mn}{2.33~{\rm{amu}}}\right)^{1/2}
\left(\frac{10~{\rm K}}{T}\right)^{1/2}
\end{eqnarray}
(CM96, eq. [10]).
For most situations of interest
($10^{-6}~{\rm{cm}} \simlt a \simlt 10^{-5}$ cm,
$\Bzci \sim 10~\mu{\rm G}$, and $T \simeq 10$ K), $\Ag \ll 1$.

We now have the necessary information to derive $\Cfac(\muci)$ in
models clouds including the effect of grain friction. Combining
equations (\ref{Cfacdefeq}), (\ref{tphiceqb}), and
(\ref{grintegraleqb})-(\ref{Ageq}), along with the results for the
ion contributions described in \S~2.1, we have, in the limit
$\Pextd \ll 1$,
\begin{mathletters}
\begin{eqnarray}
\label{Cfactoteqa}
\hspace{-1em}\Cfac(\muci) &=& \frac{\frac{1 - \muci^{2(1-k)}}{2(1-k)} + \left(\frac{\tni}{\tng}\right)_{\rm{c0}}
\left[1 - \muci\left(\frac{1 + \Ag^2}{1+\Ag^2\muci^4}\right)^{1/4}\right]}
{1 + \left(\frac{\tni}{\tng}\right)_{\rm{c0}} \left(\frac{1}{1+\Ag^2\muci^4}\right)}  \nonumber \\ \\ 
&& \hspace{15em}{\rm{if}}~k < 1, \nonumber \\
\label{Cfactoteqb}
&=& \frac{\ln\left(\frac{1}{\muci}\right) + \left(\frac{\tni}{\tng}\right)_{\rm{c0}}
\left[1 - \muci\left(\frac{1 + \Ag^2}{1+\Ag^2\muci^4}\right)^{1/4}\right]}
{1 + \left(\frac{\tni}{\tng}\right)_{\rm{c0}} \left(\frac{1}{1+\Ag^2\muci^4}\right)} \nonumber \\ \\
&& \hspace{15em} {\rm{if}}~k = 1. \nonumber
\end{eqnarray}
\end{mathletters}
Examination reveals that, in the limits
$(\tni/\tng)_{\rm{c0}}\rightarrow 0$ (dynamically unimportant dust grains),
or $\Ag \rightarrow \infty$ (grains are not
well-coupled to field lines and move with the neutrals, thereby not retarding
contraction), both equations (\ref{Cfactoteqa}) and (\ref{Cfactoteqb}) revert
to the expressions found in \S~2.1 due to the effects of ions alone,
namely, equations (\ref{Cfaceqc}) and (\ref{Cfaceqb}). 
Furthermore, for most circumstances, because
$10^{-6}~{\rm{cm}} \simlt a \simlt 10^{-5}~{\rm{cm}}$, resulting
in $\Ag \ll 1$ (see eq. [\ref{Ageqb}]), clouds with $1-\muci \ll 1$ 
have $\Cfac(\muci) \approx 1 - \muci$, just as we had found earlier.

Our derivation can be compared to numerical simulations of model clouds
that account for the dynamical effects of grains. In model B of CM94
(see their \S\S~3.1 and 3.3), $\muci = 0.26$, $\Bzci = 35~\mu$G,
$T=10$ K, $\xxici = 2.0\times 10^{-7}$, $\xgci= 7.1 \times 10^{-11}$,
and the radius of the grains $a=3.75 \times 10^{-6}$ cm. Inserting these
values into the expressions for $\tnici$ (eq. [\ref{tnieq}]), $\tngci$
(eq. [\ref{tngeq}]), and $\Ag$ (eq. [\ref{Ageqb}]), and taking
$k = 0.64$ (see Fig. $4c$ of CM94), equation (\ref{Cfactoteqa})
gives $\Cfac(\muci) = 0.86$. In actuality, CM94 model B had
$\Cfac(\muci) = \tcore/\tphici = 0.92$. Hence, our analytical
expression for $\Cfac(\muci)$ is again in very good agreement with the
results of detailed evolutionary calculations.

The core formation timescale $\tcore = \Cfac(\muci) \tphici$.
Numerically, when the collisional effects of
grains are accounted for, $\tphici$ is again given by equation
(\ref{tphicivaleq}), but multiplied by a factor
\begin{eqnarray}
\label{grainfactoreq}
1 &+& \left(\frac{\tni}{\tng}\right)_{\rm{c0}}
\frac{1}{1 + \Ag^2 \muci^4} \nonumber \\ &\approx& 
1 + 2.6 \left(\frac{T}{10~{\rm K}}\right)^{1/2}
\left(\frac{3~{\rm g}~\cc}{\rho_{\rm s}}\right)
\left(\frac{10^{-6}~{\rm{cm}}}{a}\right) \nonumber \\
&&\hspace{8em} \times
\left(\frac{10^{-7}}{\xxici}\right)\left(\frac{\chi_{\rm g}}{0.01}\right) ,
\end{eqnarray}
where equations (\ref{tnieq}) and (\ref{tngeq}) were used to evaluate
$\tnici$ and $\tngci$, and we also made the approximation
$\Ag^2 \muci^4 \ll 1$. In equation (\ref{grainfactoreq}) we have
used $\mg = 4 \pi \rho_{\rm s} a^3/3$, where $\rho_{\rm s}$ is the
density of the solid grain material, and also the relation
$\xgci = (\mn/\mg) \chi_{\rm g}$, where $\chi_{\rm g}$ is the
dust-to-gas mass ratio normalized to the canonical
value for the interstellar medium. From equations (\ref{Cfactoteqa})
and (\ref{Cfactoteqb}) it follows that clouds with $\muci > 0.5$
still have $\Cfac(\muci) \ll 1$, and therefore, $\tcore \ll \tphici$.
For the values of $\tphici$ given by equations (\ref{tphicivaleq})
and (\ref{grainfactoreq}), depending on the properties of grains
in those clouds, values of $\tcore$ in the range 1 to 5 Myr are
still likely.
\section{Relevance to Observations}
Our results reveal that dense cores may be regarded as an ensemble of
objects which are closely clustered on either side of the
subcritical-supercritical boundary, as implied by recent observations
(Crutcher 1999). The subcritical objects are still evolving due to
ambipolar diffusion, and the supercritical objects have crossed the
critical barrier and may show extended infall (e.g., the case of L1544;
CB00). This picture seems consistent with the observation that only
some cores show extended infall, and that even L1544 requires
significant magnetic forces to keep the infall speeds
to subsonic levels (CB00), given that it has little turbulent support.
However, the mean lifetime for all the cores, from statistical
studies, can still be $\simlt$ few Myr, due to the relatively short
duration of the subcritical phase at the observed densities.

Statistical studies of dense cores, usually
defined as regions in which the mean density $\nn \simgt 10^4 \, \cc$, 
since they excite the NH$_3$ $(J,K) = (1,1)$ line, reveal a lifetime 
$\sim 1$ Myr for starless cores (Lee \& Myers 1998; Jijina et al. 1999).
There are considerable uncertainties (at least of order one) in this 
determination, since the lifetimes are based upon a bootstrapping from
inferred pre-main-sequence (PMS) lifetimes; the relative statistics of
cores with embedded infrared sources to PMS sources yields a lifetime
for the embedded phase, and the relative statistics of cores with and
without embedded sources yields a lifetime for the starless phase. This
is done under the assumption that the formation times of the cores are
uncorrelated and uniformly spread in time up to the present. There is
also a question as to whether the samples are complete, as many of
them are biased toward cores that already contain stars, and regions
that have already undergone considerable star formation activity.

A cautionary note regarding lifetimes comes from the submillimeter
continuum studies of Ward-Thompson et al. (1994, 1999). In all starless
cores detected in the submillimeter regime, the inferred central density
is in the range $10^5 \, \cc - 10^6 \, \cc$, meaning that the cores
have a much shorter gravitational contraction timescale than cores
whose inferred 
central density is only $10^4 \, \cc$. In particular, at a central
density $\nnc = 3 \times 10^5 \, \cc$, the spherical free-fall time
$\tffc = (3 \pi/32 G \rho_{\rm{n,c}})^{1/2} = 6 \times 10^4$ yr, 
compared to $3 \times 10^5$ yr at a density $10^4 \, \cc$. This means
that if the submillimeter cores have a lifetime $\sim 1$ Myr, then they
live for $\sim 20 \, \tff$, making ambipolar diffusion very necessary
at this stage (see Ward-Thompson et al. 1999, who even conclude that
some ambipolar diffusion models evolve too {\it rapidly}). The
resolution to this apparent discrepancy can only come if it is
demonstrated that the submillimeter cores represent a small subset of
the ammonia cores, and high resolution studies establish that most of
the ammonia cores actually have central densities close to
$10^4 \, \cc$. A complete sample of starless cores (with high angular
resolution) is thus very necessary before drawing definitive conclusions
about lifetimes.

Another caution against preliminary conclusions about
the timescale for protostellar evolution comes from a recent study by
Kontinen et al. (2000), who measured the
abundances of various molecular species in two dense cores: the
starless, prestellar core Cr A C located in Coronae Australis, and the
Cha MMS1 core in the Chamaleon I dark cloud, which contains a class 0
source. Comparing the results of their observational study with time
dependent chemical evolution models, they argue that Cr A C exhibits
`later time chemistry' than that which is seen in the Cha MMS1 core.
Kontinen et al. suggest, as one possibility, that the reason the
Cr A C core appears to be chemically or chronologically older than the
more physically evolved Cha MMS1 core is due to the Coronae Australis
cloud having had a longer ambipolar diffusion timescale than in
Chamaleon I. Of course, based on our discussion in this paper,
differences in the ambipolar diffusion and core formation timescales
between these two parent clouds may reflect differences in the values
of $\muci$ (and, thus, $\Cfac(\muci)$), as well as disparities in
$\xxici$ and the initial central dust abundance $\xgci$ (which
could result from variations in the local cosmic-ray
ionization rate $\zeta_{\rm{CR}}$ and/or dust grain properties --- see
CM98 for a discussion on how the fractional abundances of charged
particles depend on these particular quantities), which would fix the
value of $\tphici$ (see eqs. [\ref{tphicivaleq}] and
[\ref{grainfactoreq}]) in each cloud. 

Finally, it is of interest to note that the recent Zeeman measurements
cited above seem to indicate that the mass-to-flux ratio in some
molecular clouds is comparable to the critical value for collapse.
That it is so close to the critical value likely indicates the
important role of magnetic fields in the formation, support, and
evolution of molecular clouds. Unfortunately, since a rigorous and
comprehensive theory for the formation of molecular clouds from out
of the diffuse interstellar medium is still lacking (clearly an
astrophysical problem of some importance, yet outside the purview of
our present study),
\footnote{Interestingly, ambipolar diffusion can provide a means
by which $\muci$ can be deduced for molecular clouds. As discussed
in CM96, reduction of the dust-to-gas ratio by ambipolar diffusion
(see \S~2.2, and eq. [\ref{grfraceq}] above) can be used to infer
$\muci$ from comparative observations of the
relative abundances of grains in the cores and envelopes 
of molecular clouds.}
more definitive statements
about the specific nature of $\muci$ and {\em why} it takes on values
$\sim 1$ for observed clouds cannot be made at this time.
However, our lack of knowledge as to why $\muci$ may
take on certain values does not affect our
analysis in this paper (nor our previously published models and
parameter studies, cited in \S~1), as we have formulated our result in
such a way that it can be applied to all subcritical 
($\muci \leq 1$) clouds or cloud fragments.
 
\section{Summary}   
In this paper we have reexamined the formation of protostellar cores
by ambipolar diffusion in magnetically supported interstellar molecular
clouds. Starting from basic principles, and utilizing the results of
detailed numerical simulations, we have derived an analytical relation
for the timescale for the formation of supercritical cores $\tcore$
within initially subcritical clouds as a function of $\muci$, the
initial central mass-to-flux ratio in units of the critical value for
gravitational collapse. The resulting expression is given simply by
$\tcore = \Cfac(\muci)\tphici$, where $\tphici$ is the initial central
flux-loss timescale, and the dimensionless factor $\Cfac(\muci)$ is the
{\em initial flux constant} of a cloud. Our analytical expressions
are found to be in excellent agreement with the results of detailed
numerical simulations of core formation by ambipolar diffusion in
disk-like model molecular clouds, including a recent model presented by
Ciolek \& Basu (2000) that accurately reproduces the density and
velocity structure within the L1544 protostellar core. Our analysis has
also included the dynamical effects of interstellar dust grains, which
is also found to agree well with our earlier numerical models.

We have derived simple, yet accurate expressions for $\Cfac(\muci)$,
valid for all $\muci \leq 1$. A strict upper limit
$\Cfac(\muci) \leq \ln(1/\muci)$ is obtained. Clouds with
$\muci \ll 1$ generally have $\Cfac(\muci) \sim 1$, and therefore
$\tcore \approx \tphici$, in agreement with the early analysis by
Mouschovias (1979). However, clouds with $\muci \simgt 0.5$
have $\Cfac(\muci) \ll 1$, thereby substantially reducing $\tcore$ with
respect to $\tphici$. Clouds with these values of $\muci$, which seem
to be indicated in certain regions by recent OH Zeeman observations,
have numerical values of $\tcore \simlt$ few Myr. Hence, the evidence
for comparably short timescales for protostellar collapse and evolution
(e.g., Jijina et al. 1999; Lee \& Myers 1999) is not in disagreement
with a picture in which observed dense cores are either somewhat
subcritical (and evolving towards the critical value due to ambipolar
diffusion) or are already somewhat supercritical and have started
accelerated or dynamical collapse.

\acknowledgements{GC gratefully acknowledges support from the New York
Origins of Life Center (NSCORT), and the Physics Department at RPI,
under NASA grant NAG5-7598. SB was supported by a grant from the 
Natural Sciences and Engineering Research Council of Canada. We are
grateful for helpful comments by an anonymous referee, and Steven Shore,
which served to improve our presentation.}

\newpage
%\section*{Captions to Figures}
\begin{figure}
\plotone{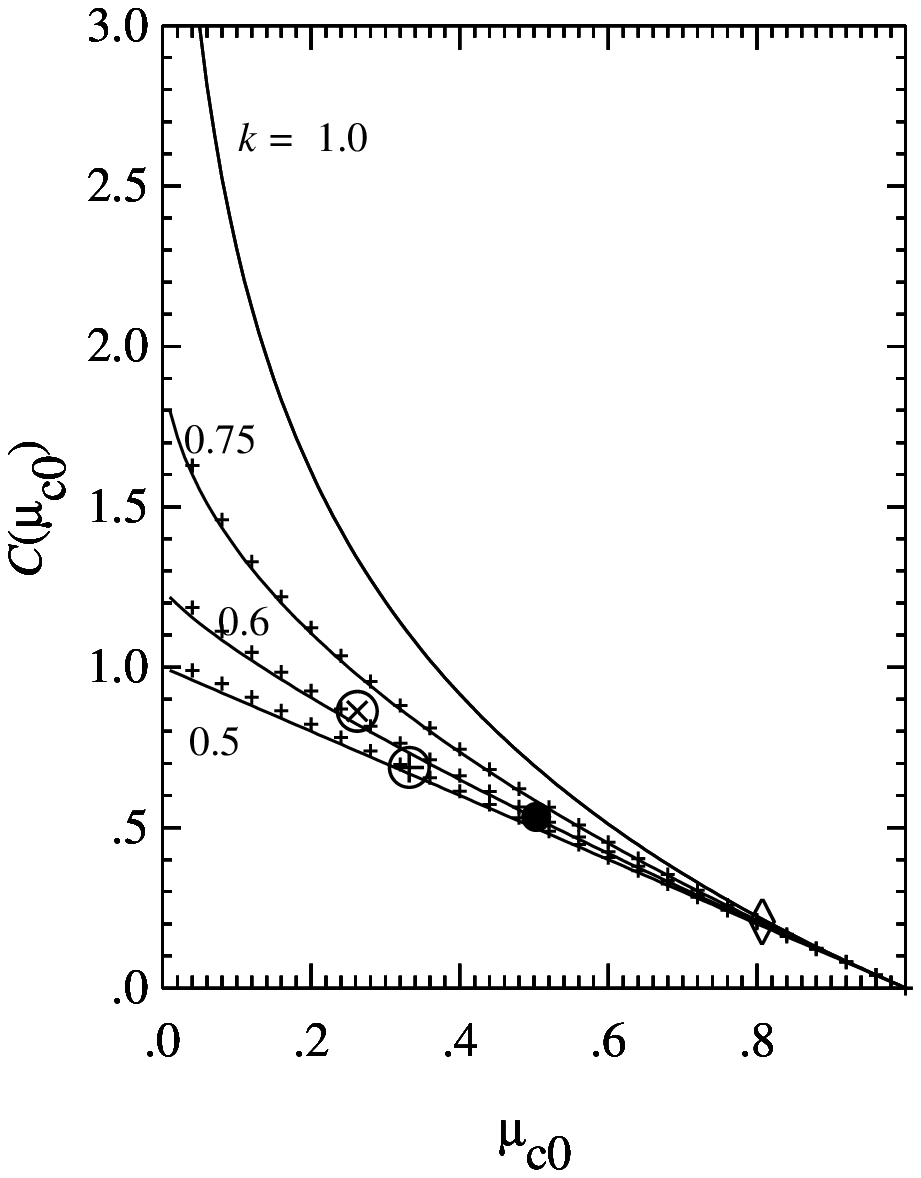}
\vspace{-30ex}
\caption{Initial flux constant $\Cfac(\muci) = \tcore/\tphici$ for
several model clouds with $k =0.5$, 0.6, 0.75, and 1.0, respectively.
{\it Crosses} represent the exact solution of equation (\ref{Cfaceqa})
for $\Pextd=0.1$ and $k =0.5$ (eq. [\ref{konehalfeq}]), 0.6, and 0.75.
{\it Solid} lines are the solutions given by equations (\ref{Cfaceqb})
or (\ref{Cfaceqc}) for
$\Pextd = 0$.
Also shown are the values found
from previously published numerical simulations: model A of CM94
(located by the center of $\otimes$), model 2 of BM94 ($\oplus$), model
5 of BM95b ($\bullet$), and the L1544 model cloud of CB00 ($\diamond$).
For most models with $\muci \ll 1$, $\Cfac(\muci) \sim 1$. Above
$\muci \simgt 0.5$, however, $\Cfac(\muci)$ is significantly less than
unity. For $1 - \muci \ll 1$, $\Cfac(\muci) \approx 1 - \muci$.}
\end{figure}
\begin{figure}
\plotone{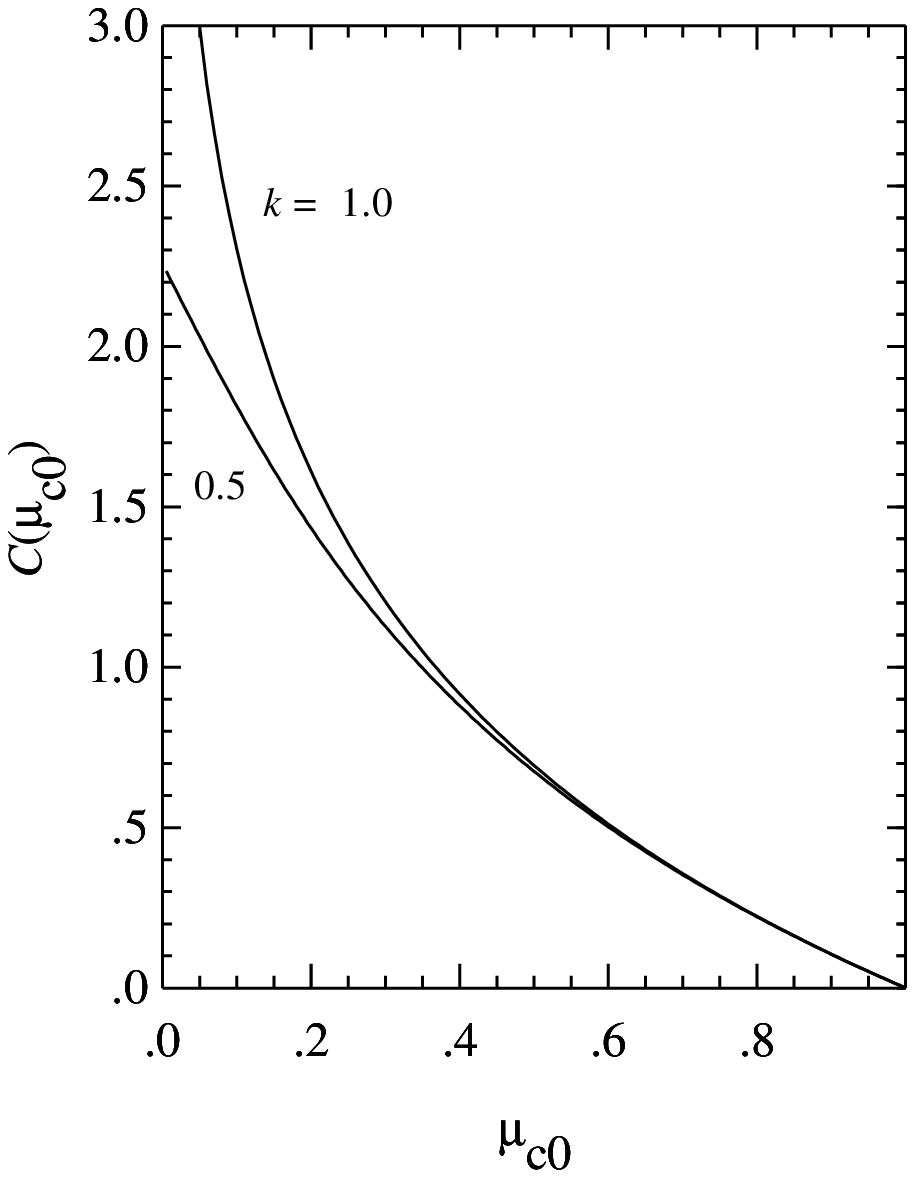}
\vspace{-30ex}
\figcaption{Same as in Figure 1, except in the limit of 
high pressure, $\Pextd = 20$, for models with $k=1.0$ and 0.5,
given by equations, (\ref{Cfaceqb}) and (\ref{konehalfeq}), 
respectively.}
\end{figure}

\end{document}